\documentclass[pre,showpacs,floatfix,twocolumn]{revtex4}
\usepackage{graphicx}
\usepackage{amssymb}
\usepackage{dcolumn}
\usepackage{bm}
\begin{document}

\title{Shear band dynamics from a mesoscopic modeling of plasticity}

\author{E. A. Jagla}

\affiliation{Centro At\'omico Bariloche, Comisi\'on Nacional de Energ\'{\i}a At\'omica, 
(8400) Bariloche, Argentina}

\begin{abstract}
The ubiquitous appearance of regions of localized deformation (shear bands) in different kinds of disordered materials under shear is studied in the context of a mesoscopic model of plasticity. 
The model may or may not include relaxational (aging) effects. In the absence of relaxational effects the model displays a monotonously increasing dependence of stress on strain-rate, and stationary shear bands do not occur. However, in start up experiments transient (although long lived) shear bands occur, that widen without bound in time. I investigate this transient effect in detail, reproducing and explaining a $t^{1/2}$ law for the thickness increase of the shear band that has been obtained in 
atomistic numerical simulations.
Relaxation produces a negative sloped region in the stress vs. strain-rate curve that stabilizes the formation of shear bands of a well defined width, which is a function of strain-rate. Simulations at very low strain-rates reveal a non-trivial stick-slip dynamics of very thin shear bands that has relevance in the study of seismic phenomena.
In addition, other non-stationary processes, such as stop-and-go, or strain-rate inversion situations display a phenomenology that matches very well the results of recent experimental studies.
\end{abstract}
\maketitle

\section{Introduction}

The mechanical response of amorphous materials is a mix between solid- and liquid-like behavior.
At low applied stresses they typically display recoverable elastic behavior, but they tend to 
flow plastically when some yield stress is overpassed. The variety of systems that fit in this 
description is very large, comprising foams and emulsions \cite{weaire0}, colloidal suspensions \cite{col}, granular matter \cite{libro1}, and metallic glasses \cite{met_glasses}.
The theory of the flowing state of these materials has developed more slowly than that of the crystalline counterpart. The reason being that for the crystalline state, there is a clear definition of a reference state onto which defects responsible for plastic flow, namely dislocations, are identified. Thus from long time ago, theories describing the energetics and dynamics of dislocations have been available. For amorphous materials there is no such reference state
and there is no an obvious definition of the kind of generalized defects that contribute to the plastic behavior.

In the last years, a theory of the plastic deformation of amorphous materials has become fashionable. This is the Shear Transformation Zone (STZ) theory \cite{stz}. It is based in early ideas by Bulatov and Argon \cite{bulatov} that plastic flow can be decomposed in 
a large amount of local plastic rearrangements that are spatially coupled through the constitutive elastic properties of the material, and that are able to generate large global deformations when summing up their individual contributions. The goal of STZ theory has been to provide rate equations for the temporal evolution of the number of STZ's as a function of time, and to show that the results are compatible with many experimental results under a variety of conditions.

One of the most ubiquitous effects in the plastic deformation of amorphous materials is strain localization. In very general terms, it refers to the coexistence of flowing and stacked, or jammed regions, within a single piece of material under the action of an externally imposed stress or deformation rate. This effect is of a great experimental importance, and it has been described in most of the realizations of amorphous materials mentioned before: complex fluids \cite{complex_fluids,coussot}, metallic glasses \cite{met_glass}, granular materials \cite{granular,fenis1,vanhecke}, and foams \cite{foams,weaire}.

In most of the experiments that have been presented so far, the localization is studied under stationary loading conditions. Yet in this situation, the sample is typically not spatially homogeneous, and the question of to what extent the observed strain localization 
is induced by a non-homogeneous geometry is a very delicate one. A particular geometry is illustrative in this respect. In the so called split-bottom experiment \cite{fenis1,fenis2}, a Couette cell formed by two concentric cylinders rotating relatively to each other is complemented by the inter-annular bottom between the two cylinders being split, with the inner (outer) part being rigidly attached to the inner (outer) cylinder.
A granular material placed in the inter-cylinder region then naturally develops a shear band that start at the position of the split line at the bottom. However, it is observed that the transition region between material moving with the inner or outer cylinder becomes progressively wider as upper layers of material are observed. This widening is in principle unbounded, indicating that in this case, a localized shear band is only induced by the particular boundary conditions by which the system is driven. 

Another interesting case of a shear band induced by a ``boundary effect" has been observed in the numerical simulations of Lennard Jones systems by Shi, Li, Kats, and Falk \cite{shi}. In this case, the use of periodic boundary conditions and homogeneous applied deformation makes the system strictly homogeneous, but the analysis of the start of deformation from a rest situation shows that shear bands appear, and become progressively wider at longer times, until they eventually involve the whole system. 
In this case, strain localization is strictly speaking a transient phenomenon, although it can last for quite a long time. 


In both of the previous experimental examples, the existence of a localized shear band is either a transient effect in time, or a localized effect in space. A natural question is thus if there are cases in which a ``persistent shear band", i.e., an infinitely long lived, finite width shear band, exists in a homogeneous spatial configuration. This question poses from the beginning difficult 
experimental challenges, since a perfectly homogeneous geometry is not realizable, and we have to rely on interpretation of non-homogeneous configurations. For instance, a cylindrical Couette cell tends to localize the strain close to the inner cylinder, because of a simple stress reinforcement mechanism due to the geometry of the system. In this case, persistent shear banding has to be searched in the detailed form of the angular velocity of the material as a function of distance from the center \cite{coussot}. 

Indirect signatures of persistent shear banding can be searched for in the bulk stress vs. strain-rate curve of the material. In fact, materials having a part of this curve with a negative slope are known to be unstable under uniform shear, and a phase separation occurs in which the system develops two regions, one of high, and another of low shear rate, very much
as in the coexistence regime of first order phase transitions. It is clear that a negative slope in the stress vs. strain-rate curve is a sufficient condition for the existence of persistent shear bands. Whether it is also necessary I think is an open question.

Recently, Manning, Langer, and Carlson \cite{manning} have shown that the results by Shi {\it et al.} \cite{shi} (in particular the time increase of shear band width) can be well fitted with results from the STZ theory, which thus is capable of describing the existence of transient shear bands. The application of the STZ theory to the case of persistent shear band has not been done in detail, although it is argued in \cite{langer} and \cite{rottler} that it requires the introduction of relaxation effects, that can be incorporated in more than a single way \cite{rottler}.

An alternative (or perhaps complementary) theoretical description of plastic flow of amorphous materials has been proposed in Ref. \cite{jagla}. In that work, I used a mesoscopic approach that (in its unrelaxed version) is conceptually similar  to the original Bulatov and Argon ideas \cite{bulatov}, and more recent extensions, as that due to Baret {\it et al.} \cite{baret,talamali}.
The system is modeled as a collection of mesoscopic pieces (see \cite{jagla} for a detailed description). Each one of these pieces is endowed with a disordered energy landscape that admits
many equilibrium configurations (and in this sense each piece contains a lot of STZ's). The dynamic state of each
element is described in terms of the two shear stress deformations $e_2$ and $e_3$ (see \cite{jagla} for detailed definitions). All pieces of the system have to fit together to satisfy elastic compatibility, and this generates an additional elastic energy in the system. Upon the external shearing, pieces of material jump between different minima of its potential energy landscape, affecting in this process the elastic state of neighbor pieces, due to the elastic compatibility conditions. With this simple model I have been able to describe for instance the flowing properties of a granular material placed in the so called spit-bottom Couette cell geometry \cite{widening}. 
However, the point that really motivated the presentation of the model in \cite{jagla}, was the possibility to introduce, in a simplified but fully consistent manner,  structural relaxation (or aging) in the system. This was introduced as a mechanism that dynamically tends to uniformize the stress within the system.
It was shown in \cite{jagla} that the introduction of this mechanism of structural relaxation generates a region of negative slope in the stress vs. strain-rate curve, and that persistent shear bands appear naturally in the model. 

In this paper I extend the use of this alternative approach to describe the plasticity of an amorphous material to show that this modeling is able to reproduce most of the phenomenology of shear bands described in the previous introduction. In particular, in the next Section, I show that in the absence of relaxation, the model describes the existence of transient 
 shear bands that widen in time as $t^{1/2}$, and I discuss the conditions for this effect to be observable. In Section III, I present results for finite relaxation in the limit of very low applied strain-rates, focusing on the appearance of stick-slip dynamics that is relevant for geological applications. In Section IV, I describe some non-stationary effects associated mainly to the rapid inversion of the stress applied to the system, showing that also in this case, the results compare very well with the outcome of recently presented experiments. Finally, in Section V, I summarize and conclude the presentation.

\section{Nucleation and widening of shear bands in the absence of relaxation}

\begin{figure}[h]
\includegraphics[width=8cm,clip=true]{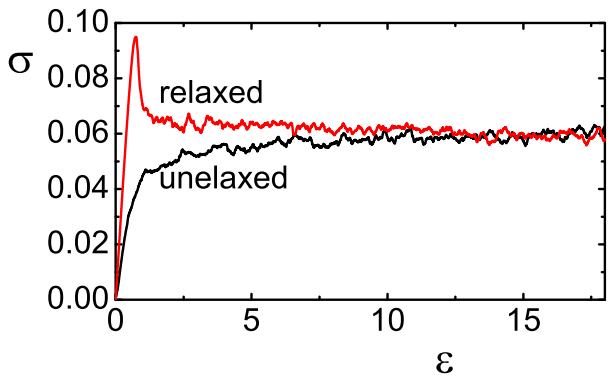}
\caption{(Color online) Stress-strain relation for an unrelaxed sample, and for a sample allowed to relax during a time $t_w=5000$ under a relaxation parameter $\lambda=0.01$. The strain is increased linearly in time, as $\varepsilon=\gamma t$, with $\gamma=2\times 10^{-4}$. System size is 128 $\times$ 128, other parameters as in Ref. \cite{jagla}.
}
\label{stress_strain1}
\end{figure}

In this section I concentrate in some properties of the model obtained in the absence of relaxation. In this case, the stress vs. strain-rate curve is monotonous (see \cite{jagla}, Fig. 3, for instance) and persistent shear bands under homogeneous spatial conditions do not occur. However, there is still room for interesting and experimentally very relevant phenomena associated to the presence of transient (although typically long lived) shear bands. 
In fact, shear bands can appear in the first stages after strain rate is applied, depending on the initial conditions of the sample.
For an amorphous material, sample preparation may involve the quench form a melt at some cooling rate. The characteristics of the sample obtained are thus dependent of this cooling rate. Typically, lower cooling rates give the system more time to reach a more relaxed configuration that make it more reluctant to shear. In fact, we will see that a relaxed initial configuration produces a yield stress peak in the stress vs. strain relation.
In the language of our model, in order to get a more relaxed initial configuration we must allow for the presence of some relaxation during the sample preparation stage. I will call this preparation time $t_w$, because it can be associated with the waiting time referred to frequently by researchers in the glass physics community.
Once the sample is prepared in this way, I set relaxation parameter $\lambda$ to zero (see Eq. 7 in \cite{jagla}), apply a train rate $\gamma$ (in such a way that $\varepsilon=\gamma t$) and observe the evolution of the system. 
This was the protocol used to obtain the stress peak in the Fig. 9 in \cite{jagla}, and I reported there that the height of that peak increases logarithmically with $t_w$. An example of the type of curves that are obtained are shown here in Fig. \ref{stress_strain1}. The applied strain in this case is of type $e_2$, corresponding to a combination of shear distortions at 45$^\circ$ of the coordinate axis. The stress reported in Fig. \ref{stress_strain1} is the one corresponding to this symmetry. Note that a very similar phenomenon had been already obtained by Bulatov and Argon \cite{bulatov}, when starting their simulation on more ordered initial configurations. 
They also showed how plastic deformation starts around the maximum of the stress vs. strain  curve, and how shear bands develop as the stress tends towards its asymptotic value. However, they did not follow in detail the widening of shear bands upon deformation.

\begin{figure}[h]
\includegraphics[width=8cm,clip=true]{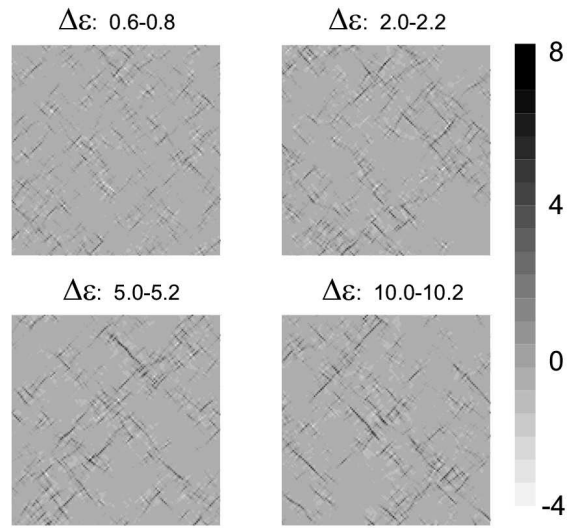}
\caption{Spatial distribution of the strain $\varepsilon\equiv e_2$ at different points of the stress vs. strain curve
of Fig. \ref{stress_strain1}, for the unrelaxed initial state. Each plot shows the deformation accumulated in the strain intervals indicated in each panel. Although there is evidence of correlated plastic rearrangements, there is no evidence of shear banding, since on average the full system participates equally of the total plastic deformation.
}
\label{spatial1}
\end{figure}

\begin{figure}[h]
\includegraphics[width=8cm,clip=true]{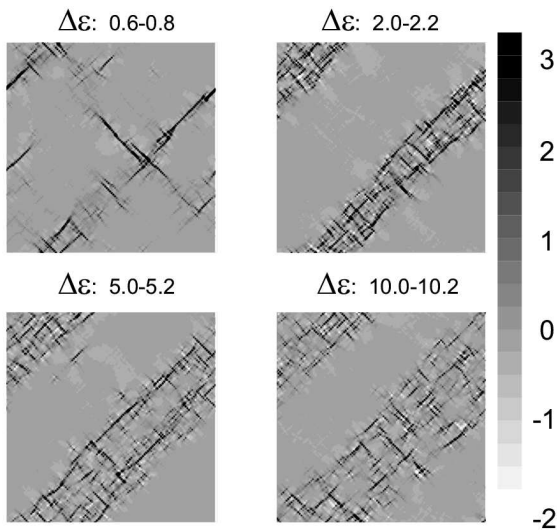}
\caption{Same as the previous figure, but now corresponding to the relaxed initial configuration Fig. \ref{stress_strain1}. Now the system has a large tendency to develop plastic regions in the form of shear bands in the first stages of the evolution. Note however, that these shear bands are not persistent, but widen progressively in time.  Eventually (not shown) the flow becomes homogeneous and qualitatively similar to that of the previous figure.
}
\label{spatial2}
\end{figure}



I discuss now the development of plastic deformation in the system, for the unrelaxed starting sample and for the relaxed one. Let us concentrate first in the unrelaxed curve in Fig. \ref{stress_strain1}. This curve does not present a stress peak. Plots of the spatial distribution of the strain deformation (Fig. \ref{spatial1}) show how the plastic deformation occurs throughout the whole sample on average. However, note that in the small strain intervals plotted in Fig. \ref{spatial1} the plastic deformation looks correlated and this might be taken as a signature of strain localization. This interpretation has been given for instance in \cite{talamali} but I think is not correct. The signatures of localization shown in Fig.  \ref{spatial1} can possibly be considered as precursors of a localization that however does not take place yet. To observe localization (yet still transient) we have to analyze the corresponding results for the relaxed initial sample. Spatial distribution of plastic deformation is seen for this case in Fig. \ref{spatial2}. 
It is observed that plastic deformation initiates close to the maximum of the stress curve, and that the spatial regions in which plastic deformation occurs are well defined and conserved during a large strain interval. 

\begin{figure}[h]
\includegraphics[width=8cm,clip=true]{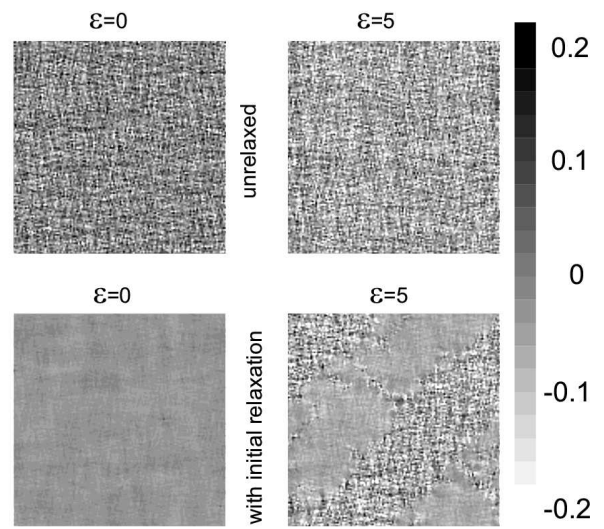}
\caption{Stress distribution at two different values of strain, for the unrelaxed (upper panels) and relaxed (lower panels) initial configuration. The effect of relaxation is clearly seen by the uniformization of the stress across the system. The relaxed state has a larger yield stress than the unrelaxed state, where  much larger stress fluctuations exist. 
}
\label{stresses}
\end{figure}

The qualitative difference between the two cases produced by the initial relaxation has to be emphasized. Remember that the relaxation mechanism acts trying to make uniform the values of the stress through the sample. 
For a relaxed starting sample, the yield stress is larger than the flowing stress. When the strain is such that the stress on the sample is reaching the yield stress value, some part of the sample fails and starts to flow. This flowing part creates a shear band\cite{notita} in the system, and now the stress decays to the flowing value {\em throughout} the sample due to a condition of mechanical equilibrium. This includes the regions that did not start to flow (thus, that retain the high yield stress of the initial sample), implying that the sample separates in a part that is flowing, and a part that is jammed and that has a yield stress higher than the actual stress. 
This image is consistent with a comparison of the spatial distribution of stress in the relaxed and unrelaxed cases  (Fig. \ref{stresses}). We can see how the relaxed starting sample presents a much smoother distribution of stresses. Once the shear band has formed, stress acquires large fluctuations within the shear band, but remains smooth outside it, i.e., the region outside the shear band maintains the high yield stress of the starting sample, whereas the band is shearing at the lower flowing value of the stress.
In these conditions there is no reason why the jammed part should start to flow, and in fact we observe the plastic deformation to be spatially localized for a rather large deformation range. However, eventually the flowing regions take over the whole sample. 
The evolution between localized an homogeneous flow seen in Fig. \ref{spatial2} occurs by a mechanism consisting of a progressive widening of the shearing regions in time. This is a very interesting effect that was observed unambiguously in atomistic simulations in Lennard Jones systems \cite{shi}. Note that in that case a peak in the stress vs. strain curve is in fact observed (originated in the way the sample was prepared) so the description of the phenomenon is the same as the one we are observing here.

It is remarkable that the time evolution of the shear band width can be predicted on very general arguments. 
Consider a sample in which an original shear band of width $w_0$ exists at $t=0$, and in which a global strain-rate $\gamma$ is being externally applied. The problem that is being considered is athermal, and for a sufficiently low strain-rate the evolution is quasistatic, implying in particular that there is no internal time scale set by the system itself. From this it is concluded that 
the velocity $v$ at which the border of the shear band advances (i.e., the rate at which its width $w$ increases) must be simply proportional to the local strain-rate inside the shear band, which is $\gamma/w$. Calling the proportionality constant $A^2/2$ for convenience, the
increase of the shear band width $w$ on time is

\begin{equation}
\frac{dw}{dt}=A^2  \gamma/2w,
\end{equation}
from which we obtain
\begin{equation}
w=\sqrt{w_0^2+A^2 \gamma t}
\label{w0}
\end{equation}
where I have set $w=w_0$ for $t=0$. The constant $A$ is a length that characterizes the material. We should expect $A$ to be a typical length in our system. The only such value in the present case is related with the typical size of the particles (or the mesh size, in the simulations), that I call $a_0$. In addition, the appearance of the strain $\gamma t$ in Eq. \ref{w} is not totally satisfactory, as the
strain scale can be easily modified in the model. This is particularly obvious in the equations of the mesoscopic model (see \cite{jagla}, footnote [17]), where it is seen that the only sensible values for the strain are those measured with respect to $\varepsilon_0$, which is the typical strain at which plastic events start to occur ($\varepsilon_0$ can be defined as the value at which the peak in the stress vs. strain ratio is observed). This means that we can conveniently set the value of $A$ as $A= B  a_0/\sqrt{\varepsilon_0}$,
where $B$ is a non-dimensional numerical constant, and thus write 
\begin{equation}
w=\sqrt{w_0^2+Ba_0^2 \frac{\varepsilon}{\varepsilon_0}}.
\label{w}
\end{equation}
The width of the shear band is thus only a function of the total deformation $\varepsilon$, independently of the rate at which the strain was applied.

We can thus make the following qualitative description of the extent of time widening of a shear band. Given a shear band with some initial width $w_0 \gtrsim a_0$, the shear strain needed to increase the width in an amount $a_0$ is $\Delta \varepsilon \sim w_0 \varepsilon_0/Ba_0$ i.e, 
an additional shear strain of order $\varepsilon_0$ inside the shear band is needed for this to thicken of the order of one particle diameter. It is clear
that this widening rate may be hardly observable for shear bands that are already much wider than the atomic size $a_0$.

\begin{figure}[h]
\includegraphics[width=8cm,clip=true]{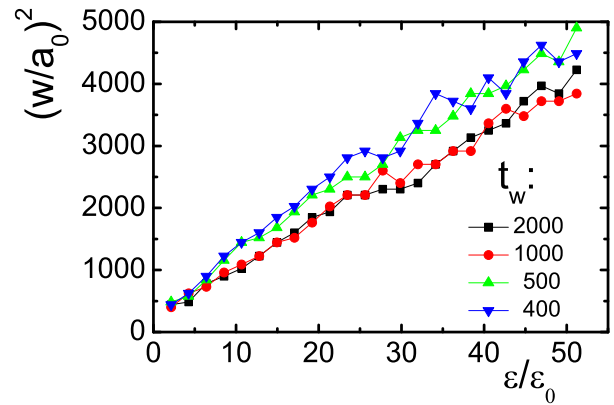}
\caption{(Color online) Evolution of the shear band width with time, in a sample that has been relaxed initially during a time $t_w$, as indicated, with a relaxation coefficient $\lambda=0.01$. After this preparation step, the strain rate applied is $\gamma=2\times 10^{-4}$. The evolution follows accurately the theoretical expectation contained in Eq. (\ref{w}). The value of the coefficient $B$ used to fit the dependence in Eq. (\ref{w}) shows a slight decrease when $t_w$ increases.
}
\label{f8}
\end{figure}

In the present mesoscopic plastic model, the shear band widening occurs precisely in the way just sketched. This widening is observed for instance in Fig. \ref{spatial2}. However, to quantitatively check the prediction contained in Eq. (\ref{w}), we face the problem that when the shear bands initially form, there may be a number of them, and the time evolution of the width is not accurately defined.
To overcome this situation, and in order to check quantitatively the widening law, I rely upon the following trick.
A shear band of a well defined width can be stabilized if the structural relaxation 
applied to generate the initial configuration does not act in a certain region in which we expect to create the shear band.
I thus act with the structural relaxation mechanism during some time $t_w$ in the absence of any strain, to set up the starting configuration, and after that, relaxation is set to zero and the strain-rate is applied.
Fig. \ref{f8} shows the time evolution of the shear band width, for different values of the time $t_w$ during which relaxation acted during the preparation stage. The linear trend in time of $w^2$ is a verification of Eq. (\ref{w}). 
The numerical value of $B$ turns out to be slightly dependent on $t_w$, corresponding to a more rapid widening for the case of lesser relaxation, i.e, a slightly more rapid widening for a less relaxed sample, which is a reasonable effect.
The order of magnitude of the values found for $B$ are around 10$^2$, which is in very good agreement with the results that can be extracted from the data in \cite{shi} Fig. 3 (using a value of $\varepsilon_0\simeq 0.03$, from \cite{shi} Fig. 2, and taking for $a_0$ in that case a typical Lennard-Jones distance parameter of the model).

\section{Thin shear bands: Stick-slip dynamics in the presence of relaxation}

I have emphasized the crucial role played by a negatively sloped stress vs. strain-rate curve in producing a persistent shear band in the system. In addition, geophysicists have known from long ago \cite{ruina,scholz} that this negatively sloped curve is
a fundamental ingredient for the seismic phenomena that occur when two tectonic plates slide one against the other. In fact, they call this phenomenon ``velocity weakening" and take it as a necessary condition for the instabilities that produce earthquakes. What is the relation between persistent shear bands occurring in velocity weakening systems and the earthquake phenomenon? In this Section I will discuss the scenario that emerges from the present model. Note that a similar analysis on the basis of the STZ theory has been made in Ref. \cite{manning2}, and that a model based on the ideas of structural relaxation presented here has been devised and applied specifically to the seismic phenomena in \cite{jagla1} and \cite{jagla2}.

It was indicated that the mesoscopic model in the presence of relaxation produces a stress vs. strain-rate curve with a region of negative slope.
The mechanical instability associated to this negative slope produces the rupture of spatial homogeneity in the system, and the appearance of persistent shear bands, as it was described in \cite{jagla}. In the presence of this shear band, the global stress vs. strain-rate behavior of the system
acquires a plateau, in which a change in strain-rate does not produce a change in stress, but it is accommodated through a change in the shear band width. This mechanism is very much in parallel with the coexistence of liquid and gas phases upon a change of volume when crossing the liquid-gas first order phase transition. In this way the mechanical stability of the system is regained, and the shear band width adjusts to the value of the externally imposed strain rate. In addition, the sliding is smooth, in the sense that stress fluctuations during sliding do not scale with system size, and are associated to the individual plastic arrangements of the system constituents. However this picture breaks down at sufficiently low strain-rates as we will see now.

In fact, if the applied strain-rate diminishes, the mechanical equilibrium conditions indicate that the width of the shear band has to decrease accordingly. But there is a limit to this possible decrease. Experimentally, a shear band cannot be thinner than a few diameters of the constituent particles in the system. In the mesoscopic model itself, the minimum width cannot be lower than a few cell sizes of the numerical mesh. In any case, when this limit is reached, the uniformly sliding solution must break down. The route that the system follows in these conditions was already indicated in \cite{jagla}: at very low strain-rates the thickness of the shear band saturates to a small fixed value, and sliding proceeds as a sequence of stick and slip events. Clear evidence of this stick-slip regime has been obtained in atomistic numerical simulations in \cite{varnik}.
One of the macroscopic manifestations of this qualitative change, is the fact that stress fluctuations in the system become progressively larger as the strain-rate diminishes. 

Before showing the concrete results in the case of the present mesoscopic model, let me point out that
I choose in this Section to work with a system that is not square, but rectangular, and in a case in which the external deformation applied corresponds to a deformation of type $e_3$. This in principle generates shear bands that run horizontally, or vertically in the system, and due to the anisotropy created by the rectangular geometry they systematically appear along the shortest length, in our case the horizontal direction. This simplifies a bit the presentation of the results, and is more convenient for a description in which a friction process (in the limit of very thin shear bands) is the main concern.

\begin{figure}[h]
\includegraphics[width=8cm,clip=true]{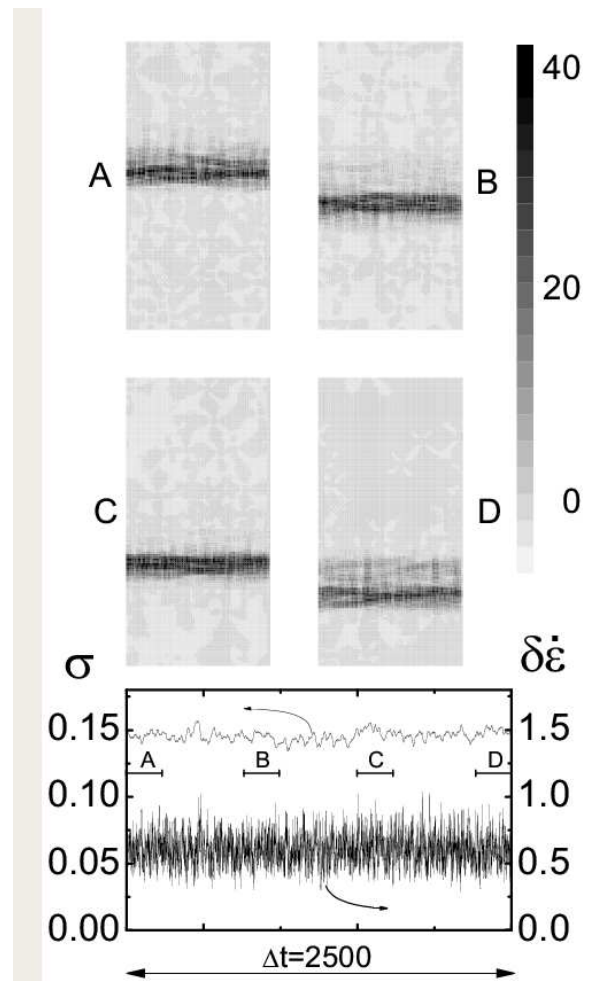}
\caption{Four snapshots of the accumulated deformation in the system for four different strain intervals (indicated in the lower panel), for a simulation with  relaxation parameter $\lambda=0.005$, and a strain rate $\gamma=0.01$. We observe the existence of a persistent shear band of a rather well defined width, which however wanders around the system as a function of time. The instantaneous stress $\sigma$ and the instantaneous spatial fluctuation of strain rate $\delta\dot \varepsilon$ are shown in the last panel. System size is 64 $\times$ 128.
}
\label{thin1}
\end{figure}

\begin{figure}[h]
\includegraphics[width=8cm,clip=true]{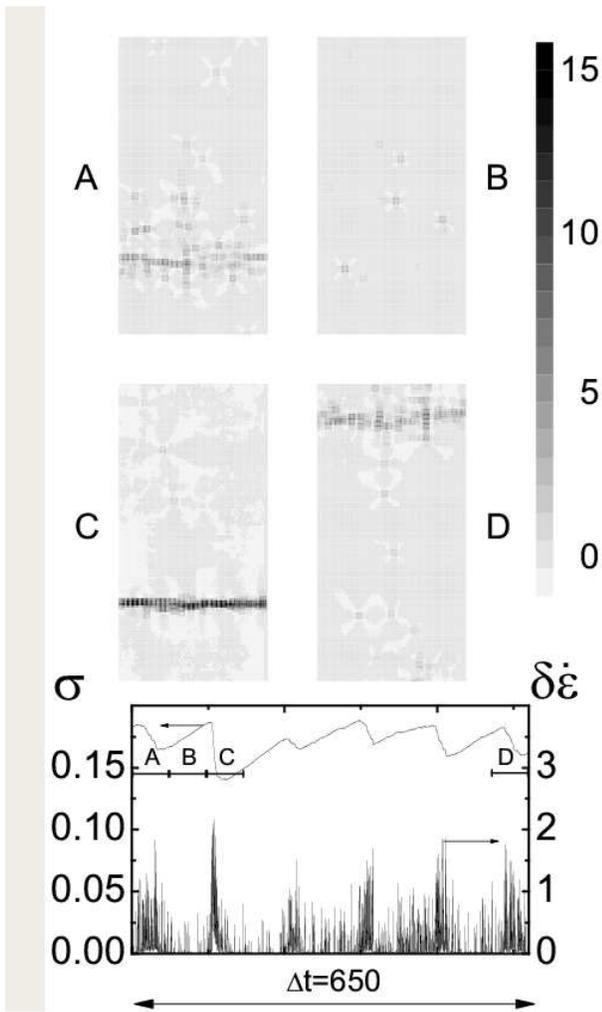}
\caption{Same as previous figure, for a lower value of the applied strain-rate, namely $\gamma=0.002$. A typical pattern of stick-slip dynamics is apparent, with periods in which the system is almost completely blocked (in the ``B" interval for instance), and others with rapid, localized deformation. Note also how these localized slips do not occur necessarily at the same spatial position, but can jump 
to a different place (compare ``C" and ``D", for instance).
}
\label{thin2}
\end{figure}

The main results of this Section are contained in Figs. \ref{thin1} and \ref{thin2}, that correspond respectively to a larger and a slower value of the applied strain-rate. 
For the case of the larger strain-rate, we observe the existence of a relatively thick shear band. The stress $\sigma$ in the system as a function of time is rather smooth in this case. I plot also in Fig. \ref{thin1} the value $\delta \dot \varepsilon$ of the spatial fluctuation of the instantaneous strain rate
For a case of uniform sliding, this quantity should be almost constant in time, and this is in fact what is observed.

For the case of a much lower strain-rate, the corresponding results are presented in Fig. \ref{thin2}. Now the stress in the system acquires a larger fluctuation, with periods of consistent increase, and others of sharp decreases, namely a typical stick-slip pattern. In fact, the intervals of linear increase of stress are correlated with periods in which strain rate fluctuations are almost absent (indicating a rigid deformation of the whole sample), whereas abrupt stress decreases correlate with large spatial strain fluctuations (corresponding to coexistence of regions that flow, and others that do not). This is also qualitatively seen in the plots of accumulated deformation in Fig. \ref{thin2}.

A remarkable characteristic to be noticed in the dynamics of the system is the fact that the average spatial location of the shear band is not constant over time, but tends to diffuse in the system. This is seen in the plots of accumulated strain in Figs. \ref{thin1} and \ref{thin2}. In the case in which the shear band is thick and the sliding uniform (Fig. \ref{thin1}), this movement is in fact a sort of diffusion of the mean position of the shear band. Runs at different strain rates indicate that the tendency to diffuse becomes stronger as the shear band becomes thinner. In the case of stick-slip motion, the shear band movement can involve a more drastic situation. In fact, it is systematically observed that subsequent slip events can occur at well separated spatial positions (compare for instance snapshots ``C" and ``D" in Fig. \ref{thin2}).
In \cite{manning2}, Daub {\it et al.} showed how strain localization (within the context of STZ theory) can induce the failing of geological gouge in a typical thickness much thinner than the gouge thickness itself. Here I am obtaining the same qualitative results (the sample here can be considered to be the gouge material), with the additional 
indication that successive failures of the gouge can occur in different spatial position. This could possibly be a relevant phenomenon to consider in the dynamic rupture process during earthquakes.

The results of this section show that the velocity weakening characteristic of the stress vs. strain-rate response in the presence of relaxation manifests very differently for the case of larger or smaller strain-rate:
when the strain-rate is relatively large, in such a way that the equilibrium width of the shear band is much larger than the mesh size (or particle size, for an experimental situation), the sliding is smooth, the expected instability of a velocity weakening system being cured in this case by the possibility to adjust the width of the shear band to the externally imposed strain-rate. When the imposed strain-rate becomes very small, in such a way that the equilibrium width of the shear band would nominally be smaller that the mesh size, the system enters a stick-slip regime as usually expected for systems with velocity weakening. 
The transition between the two regimes is seen not to be abrupt, instead a smooth crossover is observed.

\section{Application to shear reversal experiments}

In the last time there has been some interest in analyzing the response of amorphous or granular materials to non-stationary stressing conditions. One particular case corresponds to an externally imposed flow condition that is stopped and then re-initiated, in the same or in the opposite direction. 
Asymmetries between the two cases are clearly detected experimentally \cite{granular,losert}. If the flow is stopped and re-initiated in the same direction, stress recovers immediately to the value prior to flow stop. However, if flow is started in the opposite direction, there is a typical strain interval in which the stress increases smoothly before reaching the steady state value. 
Here I will show that these results are obtained in the present model of plasticity, and also give a very simple description in terms of a mean field analysis. In all this section structural relaxation is set to zero, this means in particular that only transient strain localization phenomena can occur, but no persistent shear banding.

\begin{figure}[h]
\includegraphics[width=8cm,clip=true]{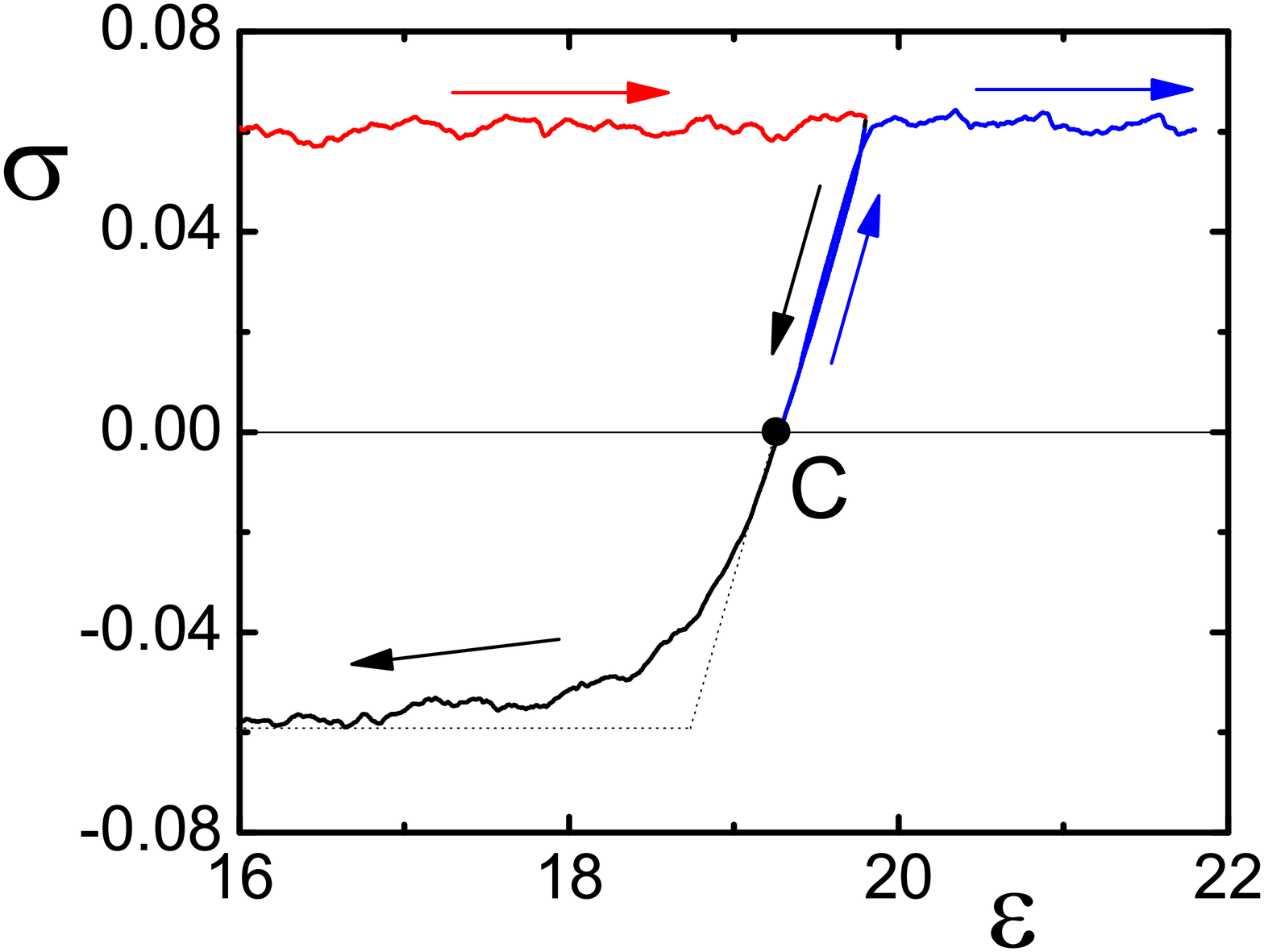}
\caption{(Color online) Stress in the system in the absence of relaxation and under spatially homogeneous flow, upon inversion of the strain-rate. For a reversion to a state of zero stress, and reloading in the same direction, the recovery of the original configuration is instantaneous beyond an elastic unloading-reloading effect. When the flow is reverted during a long time the stress reaches the negative of the original value following a smooth evolution (compared with an immediate recovery, as indicated by the dotted line), indicating that the system takes some finite strain interval to accommodate to the new flowing conditions.
}
\label{inversion1}
\end{figure}

I will first consider a spatially homogeneous situation of uniform flow under a constant applied strain-rate.
Let us suppose that in these conditions the applied strain-rate is suddenly reversed during some time. Upon this inversion the stress in the system has to revert eventually to minus its original value (the system is symmetric upon a change of sign of strain). However, the response of the system is not instantaneous. In Fig. \ref{inversion1} we see the evolution of stress upon this process, in two different cases. In the first case strain-rate is inverted only up to the point C where stress becomes zero, and then re-initiated again in the original direction.
In this case the system responds elastically, the strain reduces almost linearly with the decreasing strain, and recovers also linearly up to the original strain value when strain is increased again. Once at the original point, upon a further strain increase the stress continues at the same constant level it had prior to inversion. This kind of behavior can reasonably be termed ``immediate recovery" upon reloading.
The second case corresponds to a permanent flow reversion. We see in this case how the stress reaches the asymptotic value (equal to minus the original one) rather smoothly, instead of what would be an immediate response, as indicated by the dotted line in Fig. \ref{inversion1}. This kind of results, that has been interpreted in \cite{losert} in terms of ``chain forces" that are present in the system and that take some time to be rebuilt upon stress inversion, can be qualitatively understood using 
a simplified mean field description of the model studied here, that I will now present.

\begin{figure}[h]
\includegraphics[width=8cm,clip=true]{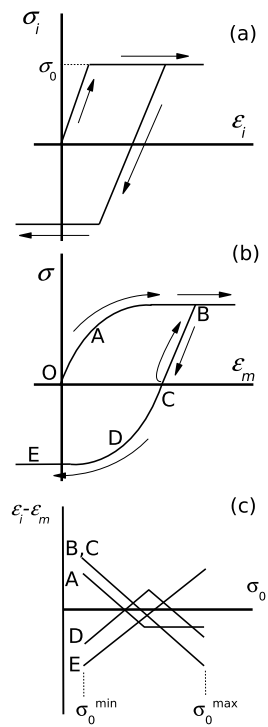}
\caption{A mean field interpretation of the flow inversion experiments. A collection of individual plastic elements with the stress vs. strain characteristic shown in (a) is considered. The values of $\sigma_0$ are broadly distributed for different elements. In (b) we see the average stress as a function of the average strain for a collection of elements as in (a). The stress increase is smooth due to the broad distribution of $\sigma_0$ in (a). In (c) the distribution of strains of the different elements as a function of its flowing stress value is plotted for different points in (b). Note in particular how the distribution corresponding to point C keeps a memory of the flowing state prior to flow stop: elements with lower (larger) $\sigma_0$ have strain values larger (lower) than the average. This justifies the asymmetry of the response upon resuming flowing in the original, or in the opposite direction.
}
\label{mf}
\end{figure}

The mesoscopic plastic description of the present paper assumes a disordered energy landscape,
independently chosen for every point of the system in the plane of the two
shear distortions that are called $e_2$ and $e_3$. 
Let me now think in a simplified description in which we concentrate in a single type of distortion (I call it simply $\varepsilon$, for definition)
and assume each elemental piece of the system behaves as an ideal plastic element as indicated in Fig. \ref{mf}(a), i.e., the stress at each element increases linearly upon strain increase, until a maximum stress $\sigma_0$ is reached, in which case a further increase of $\varepsilon$ does not increase the value of $\sigma$. Upon strain-rate inversion, stress goes down linearly to $-\sigma_0$, and remains there. Different spatial positions are assumed to have different values of $\sigma_0$. 

Driving the system (through the mean value of $\varepsilon$, namely $\varepsilon_m=\overline {\varepsilon_i}$), produces a stress that is obtained as the spatial average of the local values, i.e, $\sigma=\overline {\sigma_i}$. The typical form of $\sigma$ vs $\varepsilon_m$ curve can be seen in Fig. \ref{mf}(b). The smooth increase of $\sigma$ is due to the broad distribution of values of $\sigma_0$. If the driving is stopped (at point B of Fig. \ref{mf}(b)), and strain reduced to the point of zero stress (C), the system retains a memory about the sense in which it was flowing, that produces the asymmetric response upon resuming deformation in the same or opposite direction. This is most clearly seen in the plots in Fig. \ref{mf}(c) where the value of deformation of each element with respect to the mean value is plotted against its flowing stress value $\sigma_0$ (for simplicity, here I assume a uniform distribution of the values of $\sigma_0$ between $\sigma_0^{min}$ and $\sigma_0^{max}$). For a flowing state in one direction, the elements with larger values of $\sigma_0$ are lagged with respect to those with the smaller values, and this produces the anisotropy. When the flow is reverted, there is a transient in which this situation has to be rebuilt for the new flowing direction. There would not be any asymmetry if all elements have the same $\sigma_0$ value. This justifies the asymmetry observed in the simulations using the mesoscopic plasticity model, as in Fig. \ref{inversion1}. 

\begin{figure}[h]
\includegraphics[width=8cm,clip=true]{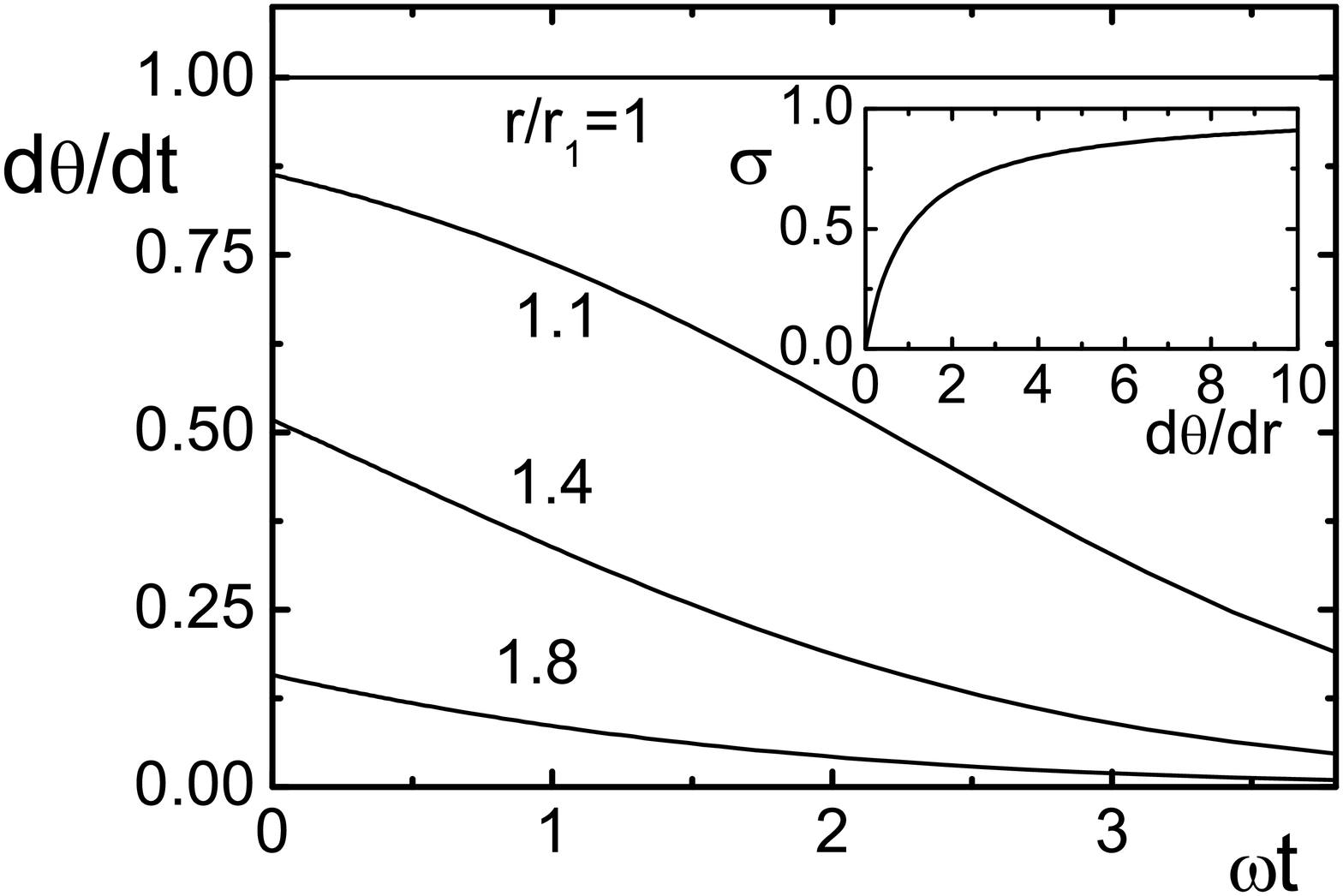}
\caption{Angular velocity as a function of time at different radial position in Couette cell experiment, after inversion of rotation direction. The cylinder radios are $r_1$, and $r_2=2r_1$. The curves were obtained assuming a local stress vs. strain that is plotted in the inset (following the law $\sigma(x)\propto x/(x+1)$), and requiring mechanical equilibrium to hold. Note that as $t\to \infty$, the shear band becomes singularly localized at the inner cylinder.
}
\label{couette}
\end{figure}

Another observation that has been made concerning flow inversion experiments, is the fact that regions that are jammed during forward flow, become more mobile during a transient period after inversion \cite{granular,losert}. For instance, in \cite{losert} the case of the flow of a granular material in a Couette cell was analyzed. In a stationary regime, the angular velocity of the material is maximum at the inner cylinder and decays rapidly with radial distance (the external cylinder is at rest). Upon flow inversion, and during some time, the decay of angular velocity with radial distance is much slower, and regions that were at rest participate of the flow, until the stationary situation sets in. Again this phenomenology is consistent with the present findings if we admit that we can represent the system as a collection of pieces having a stress vs. strain ratio like that in Fig. \ref{mf}(b), and that are located at different distances $r$ from the center. 
We can analyze the response in the following way. 
Consider that the inner and outer cylinder are located at radial distances $r_1$ and $r_2$, and the angular velocity is fixed at them as $\omega$, and $0$, respectively.
Mechanical equilibrium implies that the product $r\sigma$ has to be constant across the system at every time, i.e, $r\sigma=C(t)$. The value of $\sigma$ is a function of the local strain, i.e., of $d\theta/dr$, so the instantaneous value of $\theta (r)$ can be obtained as
\begin{equation}
\theta(r)=\int_r^{r_2} \sigma^{-1}(C/r)dr
\end{equation}
where $\sigma^{-1}$ is the inverse function of $\sigma$. The time dependence of the $C$ value is implicitly 
determined by the fact that the inner cylinder is rotated at an angular velocity $\omega$, i.e.,
\begin{equation}
\omega t=\int_{r_1}^{r_2} \sigma^{-1}(C/r)dr
\end{equation}
In Fig. \ref{couette} I show the result of solving the previous two equations for a particular form of the function $\sigma$ (namely $\sigma(x)\equiv x/(1+x)$), and we see how a transient of enhanced mobility is observed upon flow reversion. After this transient, velocity of all points excepts the inner cylinder decay to zero, as in this description the equilibrium shear band is singularly localized in the point of maximum stress \cite{otranota}.

\section{Conclusions}

In this paper I have presented results that indicate that a mesoscopic theory of plasticity complemented, if necessary, with the inclusion of relaxational (aging) terms is able to reproduce much of the available phenomenology associated with plasticity in different kinds of amorphous materials. A summary of the different possibilities that have been described is the following. 
First of all, it has been argued that a persistent shear band (i.e., one whose formation is not associated to inhomogeneities of the experimental set up, and that persists indefinitely with a well defined width) can only exist for materials that have a part of the stress vs. strain-rate curve with a negative slope. In my modeling, this condition is fulfilled only in the presence of structural relaxation. Under this conditions a shear band forms in the system satisfying formally a sort of ``Maxwell construction" as occurs in coexistence regions of first order phase transitions. In the present case the stress plays the role of the pressure (thus remaining constant at coexistence), and the shear band width plays the role of the relative fraction of the two coexistent phases. In this situation sliding is smooth. However, if strain-rate is so slow that equilibrium shear band width would be smaller than some atomic length in the model, the sliding becomes of the stick-slip type. In this limit the model provides an interesting conceptual tool to study the characteristics of seismic phenomena \cite{jagla1,jagla2}.

In all cases in which structural relaxation is absent and the stress vs. strain-rate curve is monotonously increasing, a shear band can exist at most as a transient phenomenon, or induced by particular inhomogeneous boundary conditions. I have discussed cases of these two possibilities: in one case, namely the split bottom Couette flow configuration, the shear is stationary in time, and the shear band is induced to appear at the split line in the bottom of the container. The width of the shear band increases as layers progressively away of the bottom are observed. The second case is a perfectly homogeneous spatial situation (obtained by the use of periodic boundary conditions) and a flow that is initiated at some fixed time. There are two possibilities in this case. If the initial configuration of the system was unrelaxed, regions of plastic deformation appear all across the sample, the stress vs. strain curve grows continuously to the asymptotic value, and no sign of spatial strain localization appears. However, in the case in which the starting sample is relaxed, 
the stress vs. strain curve develops a peak (the yield stress), plastic deformation is localized in space, forming a shear band, and starts to occur close to strains where the stress develops the peak. This shear band however widens in time with a $t^{1/2}$ law that was explained in terms of the athermal and quasistatic nature of the process. Eventually, for very long times, the shear band broadens sufficiently to cover the whole system, and deformation becomes spatially  uniform. 
If in this homogeneous and stationary state, external deformation is stopped and then re-initiated in the same original direction, the stress recovers its original value almost instantaneously (beyond some elastic unloading-reloading stage). However, if deformation is re-initiated in the opposite direction, there is a reacommodation stage during which the stress in the system grows smoothly before reaching its asymptotic value. This behavior was interpreted in term of a simplified version of the mesoscopic plasticity model.

On a general perspective, I think that the present kind of modeling is an interesting alternative to other more microscopic approaches (as for instance the STZ theory) to study the different possibilities of plastic deformation of amorphous materials. 
On one hand, the results in the absence of relaxation have been shown to be compatible with different equilibrium, and nonequilibrium phenomena, as the shear band widening described in \cite{shi}, and the stress inversion experiments in \cite{losert}.
On the other hand, the inclusion of relaxation terms opens new routes in the study of physical properties of systems with velocity weakening features, in particular related to the seismic phenomena \cite{jagla1,jagla2}.

\section{Acknowledgments} 

Fruitful discussions with M. Falk and W. Losert are greatly acknowledged.
This research was financially supported by Consejo Nacional de Investigaciones Cient\'{\i}ficas y T\'ecnicas (CONICET), Argentina. Partial support from
grants PIP/112-2009-0100051 (CONICET, Argentina) and PICT 32859/2005 (ANPCyT, Argentina) is also acknowledged.


\begin{thebibliography}{15}
\bibitem{notita} The appearance of one or more shear bands in the system in this case seems to depend on particular
conditions associated to the fluctuations of stresses across the whole sample. The expectation is that a larger stress rate will produce on average more shear bands.
\bibitem{otranota}Note that in the experiments in \cite{losert} an equilibrium non-singular profile is obtained. This is associated with a material that has a stress that increases with strain-rate. In the simple mean field description given here stress does not depend on strain-rate, and the shear becomes singularly localized at the inner cylinder for long times.
\end{thebibliography}
\end{document}